\documentclass{article}
\usepackage{spconf,amsmath,graphicx}
\usepackage{multirow}
\usepackage{amssymb}
\usepackage{graphicx}
\usepackage{algorithm}
\usepackage{amssymb}
\usepackage{times,amsmath,epsfig}
\usepackage{amsmath,graphicx,latexsym}
\usepackage{algorithm,algorithmic}
\usepackage{amssymb}
\usepackage{url}
\usepackage{times,amsmath,epsfig,changebar}
\usepackage[tight]{subfigure}

\newcommand{\comment}[1]{}

\title{Privacy Preserving and Collusion Resistant Energy Sharing}
\name{Yuan Hong, Han Wang, Shangyu Xie, Bingyu Liu}

\address{Department of Computer Science\\Illinois Institute of Technology\\10 W 31st Street, Chicago, IL 60616\\yuan.hong@iit.edu}

\begin{document}
%\ninept
%
\maketitle
\begin{abstract}

Energy has been increasingly generated or collected by different entities on the power grid (e.g., universities, hospitals and householdes) via solar panels, wind turbines or local generators in the past decade. With local energy, such electricity consumers can be considered as ``microgrids'' which can simulataneously generate and consume energy. Some microgrids may have excessive energy that can be shared to other power consumers on the grid. To this end,  all the entities have to share their local private information (e.g., their local demand, local supply and power quality data) to each other or a third-party to find and implement the optimal energy sharing solution. However, such process is constrained by privacy concerns raised by the microgrids. In this paper, we propose a privacy preserving scheme for all the microgrids which can securely implement their energy sharing against both semi-honest and colluding adversaries. The proposed approach includes two secure communication protocols that can ensure quantified privacy leakage and handle collusions.

\end{abstract}
%
%\begin{keywords}
%Smart Grid, Privacy, Security
%\end{keywords}
%

\section{Introduction}
\label{sec:intro}

Energy has been increasingly generated or collected by different entities on the power grid (e.g., universities, hospitals and householdes) via solar panels, wind turbines or local generators in the past decade. With local energy, such electricity consumers can be considered as ``microgrids'' which can simulataneously generate and consume energy \cite{SaadHPB12,ArboleyaGCFMSBP15}. More recently, the research on cooperation among entities on the power grid (e.g., microgrids) has attracted great interests in both industry and academia \cite{SaadHPB12}. For instance, microgrids can share their local energy to improve the efficiency and resilience of power supply \cite{HongIJER15}. 

Specifically, microgrids can transmit their excessive energy to the microgrids close to them. In the cooperation, all the participating microgrids jointly seek an energy transmission assignment that minimizes the global energy loss during transmission. However, to this end, all the microgrids should disclose their local information (e.g., local supply, local demand, and power quality for transmission) to each other or a third party. Then, the data recipient (which is a microgrid or a third party) formulates an optimization problem by denoting the amount of energy transmitted from $M_i$ to $M_j$ as $x_{ij}$ and determining the objective function as well as the constraints.

Disclosing such local information to each other or a third party would compromise the corresponding microgrid's local information. To tackle the privacy concerns, the proposed approach in \cite{HongIJER15} efficiently transforms the shares of the optimization problem to a privacy-complaint format and enables any party to solve the problem. However, the algorithms in \cite{HongIJER15} pursue high efficiency but cannot quantify the privacy leakage in the protocol. In this paper, we extend the transformation and optimal solution reconstruction to two secure communication protocols in which privacy leakage can be quantified and bounded. In the meanwhile, we give formal security/privacy analysis for the protocols and identify that our proposed secure communication protocols can prevent additional information leakage against the potential collusion among microgrids while executing the protocols. Finally, we present some experimental results to demonstrate the effectiveness and efficiency of our approach.

\section{Related Work}
\label{sec:related}
In smart grid infrastructure, privacy concerns were recently raised in the fine-grained smart meter readings, which is frequently reported to the utility \cite{SankarRMP13,HongTIFS17,SGSBook}.  To prevent information leakage in smart metering, three different categories of privacy preserving schemes were proposed in the past few years. The first category of techniques built cryptographic protocols to directly aggregate or analyze such meter readings without sharing the raw data. For instance, Rottondi et al. \cite{RottondiVC13} proposed a privacy preserving infrastructure based on cryptographic primitives to enable utilities and data consumers to collect and aggregate metering data. The second category of techniques obfuscate the meter readings to prevent adversaries from learning the status of the appliances at different times. For instance, Hong et al. \cite{HongTIFS17} defined a privacy notion to quantitatively bound the information leakage in smart meter readings, and proposed streaming algorithms for converting the readings with guaranteed output utility. Finally, the third category of techniques utilize renewable energy sources like batteries to hide the actual load of different households, which can be found in \cite{McLaughlinMA11}, \cite{YangLQQMM12}, etc.

Furthermore, energy sharing problem among microgrids \cite{SaadHPB12,ZhuHSSIMMS13} has been recently studied -- locally generated energy can be shared among homes due to the mismatch between generation harvesting and consumption time in microgrids. Zhu et al. \cite{ZhuHSSIMMS13} developed an energy sharing approach to determine which homes should share energy, and when to minimize system-wide efficiency loss. Zhu et al.  \cite{ZhuXPTG11} also proposed a secure energy routing approach to renewable energy sharing against security attacks such as spoofed routing signaling and fabricated routing messages. Also, some game theoretical models \cite{SaadHPB12,MaityR10,DuanD09} were proposed to mitigate the risks of self-interested behaviors in the energy sharing/exchange. So far, Hong et al. \cite{HongIJER15} is the only work that resolves the privacy issues in energy sharing/exchange. The proposed scheme can provide some ad-hoc privacy guarantee based on matrix multiplication. Instead, we extend the approach in \cite{HongIJER15} to ensure provable security.

\section{Preliminaries}
\label{sec:pre}
In this section, we briefly summarize the problem formulation, transformation and solution reconstruction in \cite{HongIJER15}. Note that the formulations of three optimization problems in \cite{HongIJER15} are similar, which can be securely transformed and solved using the same secure communication problem. Thus, we only focus on the basic formulation. 

\subsection{Problem Formulation}

Given $n$ microgrids $M_1,\dots, M_i$, the demand and supply of $M_i$ at time $t$ is denoted as $D_i(t)$ and $S_i(t)$, respectively. Then, given $x_{ij}$ as the amount of energy transmitted from $M_i$ to $M_j$, the optimization (LP) problem to minimize the overall energy delivery loss in the sharing is formulated as follows.

\begin{equation}
\small
\begin{gathered}
\min: \sum_{i=1}^{n}\sum_{j=1}^n \theta_{ij}x_{ij}\\
s.t.{\centering
	\begin{cases}
	\forall i\in[1,n], \sum_{j=1}^n(1-\theta_{ji})
	x_{ji}-\sum_{j=1}^nx_{ij}+S_i(t)\geq D_i(t)\\
	\forall i\in[1,n], \sum_{j=1}^nx_{ij} \leq S_i(t)\\
	\forall i\in [1,n], \forall j\in[1,n], x_{ij}\geq 0
	\end{cases}}
\end{gathered}\label{eq:meet}
\end{equation}

where $\theta_{ij}$ represents the energy loss rate for transmission between $M_i$ and $M_j$, which is determined by the distance between them on the power transmission network and the power quality data, such as voltage and current. $D_i(t)$, $S_i(t)$ and $\theta_{ij}$ are privately held by microgrid $M_i$. The general form of Equation \ref{eq:meet} can be derived as below:

\begin{equation}
\begin{gathered}
\min ~~~~~\vec{c_1}^T\vec{x_1}+\vec{c_2}^T\vec{x_2}+\dots+\vec{c_n}^T\vec{x_n}\\
s.t.{\centering
	\begin{cases}
	\begin{array}{ccc}
	A_1\vec{x_1}+& \ldots &+ A_n\vec{x_n} \\
	B_1\vec{x_1}&  & \\
	& \ddots&\\
	& & B_n\vec{x_n}\\
	%\vec{z_1}, & \dots & ,\vec{z_n}
	\end{array}\begin{array}{c}
	\leq\\
	\leq\\
	\vdots\\
	\leq\\
	%\geq
	\end{array}
	\begin{array}{c}
	\vec{b_0}\\
	\vec{b_1}\\
	\vdots\\
	\vec{b_n}\\
	\end{array}
	\end{cases}}
\end{gathered}
\label{eq:std}
\end{equation}

where $\vec{x_i}$ represents $M_i$'s variables $(x_{i1},\dots, x_{in})$, which is privately held by $M_i$. Matrices/vectors $A_i$, $B_i$, $\vec{c_i}^T$ and $\vec{b_i}$ are $M_i$'s private inputs in the LP problem. 

\subsection{Transformation}

The above LP problem is heterogeneously partitioned into $n$ shares -- global constraints are co-held by all the parties (vertically partitioned \cite{HongJCS12,HongDBSEC14,HongThesis}) while each constraint belongs to only one party (horizontally partitioned \cite{HongOpt13,HongJIS14,HongTDSC15}). To ensure privacy protection in solving and realizing the above problem, a transformation-based approach \cite{HongIJER15} was proposed:

\begin{align}
\forall i\in[1,n], A_i&\longrightarrow A_iQ_i\nonumber\\
\forall i\in[1,n], B_i&\longrightarrow B_iQ_i\nonumber\\
\forall i\in[1,n], \vec{c_i}^T&\longrightarrow \vec{c_i}^TQ_i\nonumber\\
\forall i\in[1,n], \vec{x_i}&\longrightarrow \vec{y_i}
\label{eq:trans}  
\end{align}

where each party $M_i$ locally post-multiplies its shares (i.e., $A_i$, $B_i$ and $\vec{c_i}^T$) in the LP problem by an $n\times n$ random nonnegative monomial matrix $Q_i$ \cite{HongJCS12} which is privately generated by itself, and variables in the new problem $\forall \vec{y_i}$ correspond to $\forall \vec{x_i}$. Then, $\forall i\in[1,n], A_iQ_i, B_iQ_i, \vec{c_i}^TQ_i$ can be disclosed to other parties.

Note that the righthand side values $\vec{b_0},\vec{b_1},\dots,\vec{b_n}$ are also transformed to random numbers in \cite{HongIJER15}, and we still keep such transformation. Thus, we will focus on the security/privacy improvement on the transformation in Equation \ref{eq:trans}.

\subsection{Reconstruction}

In \cite{HongIJER15}, after solving the transformed problem to obtain the optimal solution $\forall \vec{y_i}^*$, the solver (any party or an external party, e.g., the cloud) distributes the solution shares to the corresponding parties. Then, the optimal solution of the original problem $\forall i\in[1,n], \vec{x_i}^*$ can be locally reconstructed as: $\vec{x_i}^*=Q_i\vec{y_i}^*$ \cite{HongIJER15,HongDBsec,HongJCS12}. The solver and other parties cannot learn the details of $\vec{x_i}^*$, $A_i$, $B_i$ since $Q_i$ is unknown to them.

\section{Extended Transformation}
\label{sec:ext}
With the transformation in \cite{HongIJER15}, each party's share of problem cannot be learnt by other untrusted parties, even if the transformed shares are disclosed to them. However, the information leakage in the communication protocol cannot be quantified. We now extend it to a more secured transformation based on Homomorphic cryptosystem (e.g., Paillier \cite{Paillier99}).\footnotemark[1]

\footnotetext[1]{Homomorphic cryptosystem is a semantically-secure public key encryption with an additional property to generate the ciphertext of an arithemetic operation between two plaintexts by other operations between their individual ciphertexts. For instance, two encryptions $E(A)$ and $E(B)$, there exists operations *, such that $E(A*B)=E(A)*E(B)$ where * is either addition or mutiplication (in some abelian group).}

\subsection{Overview}

The basic idea of the extended transformation is described as follows. For any party $M_i$'s shares in the LP problem $A_i$, $B_i$ and $\vec{c_i}^T$, we let all the parties jointly transform such shares (via Homomorphic Encryption) in sequence -- while transforming $M_i$'s shares, party $M_j$ locally generates a new random nonnegative monomial matrix $\forall j\in[1,n], Q_{ij}$, and post-multiplies it to each of the three transformed shares (by the previous party). In case that $j=i$ holds, $M_i$ post-multiplies its own shares by its own matrix $Q_{ii}$. Similarly, all the parties jointly reconstruct every share of the optimal solution $\vec{y}$ by pre-multiplying their matrices in a reverse order (also via Homomorphic Encryption). 

\subsection{Extended Secure Transformation}

Without loss of generality, we let an external party $P$ (e.g., the cloud) solve the transformed problem. In the extended secure transformation protocol, $P$ generates the public/private key pair $(pk, sk)$, and distributes the public key $pk$ to $M_1,\dots, M_n$. Since the transformation for $A_i, B_i$ and $\vec{c_i}^T$ are identical \cite{HongIJER15}, we can take $A_i$ as an example to illustrate our secure transformation protocol in Algorithm \ref{algm:tran}.

\begin{algorithm}[!h]
	\small
	\begin{algorithmic}[1]
		\renewcommand{\algorithmicrequire}{\textbf{Input:}}
		\renewcommand{\algorithmicensure}{\textbf{Output:}}

		\FOR{$i \in [1,n]$} 
		\STATE $M_i$ randomly generates $Q_{ii}$ 
		
		\STATE $M_i$ encrypts $A_iQ_{ii}$ with $pk$ to generate $E=Enc_{pk}(A_iQ_{ii})$, and sends $E$ to the next party $M_j$ in Line 4
		
		\FOR{$\forall j\in [1,n], j\ne i, M_j$}
		
		\STATE $M_j$ randomly generates $Q_{ij}$

		\STATE $M_j$ updates $E$ with $Q_{ij}$ (Line 7-9: $E_{ab}$ denotes the entry at row $a$ and column $b$ in $E$, and $(Q_{ij})_{kb}$ denotes the entry at row $k$ and column $b$ in $Q_{ij}$)
			
		\FOR {each row $a$ of $E$ and each column $b$ of $Q_{ij}$}
		\STATE $M_j$ computes $E_{ab} \leftarrow \prod^{n}_{k=1} E_{ak}^{(Q_{ij})_{kb}}$
		\ENDFOR
		
		\STATE $M_j$ sends the updated $E$ to the next party
		
		\ENDFOR
		
		\STATE the last party sends $E$ to the solver $P$
		
		\STATE $P$ decrypts $E$ to obtain: $A_iQ_{ii}\prod_{j=1,j\ne i}^nQ_{ij}$
		
		\ENDFOR
	
\end{algorithmic}
	\caption{Extended Secure Transformation}\label{algm:tran}
\end{algorithm}

After decrypting all the ciphertexts, solver $P$ can forumate a new LP problem with the transformed shares:

\begin{align}
\forall i\in[1,n], A_i&\longrightarrow A_iQ_{ii}\prod_{j=1,j\ne i}^nQ_{ij}\nonumber\\
\forall i\in[1,n], B_i&\longrightarrow B_iQ_{ii}\prod_{j=1,j\ne i}^nQ_{ij}\nonumber\\
\forall i\in[1,n], \vec{c_i}^T&\longrightarrow \vec{c_i}^TQ_{ii}\prod_{j=1,j\ne i}^nQ_{ij}\nonumber\\
\forall i\in[1,n], \vec{x_i}&\longrightarrow \vec{y_i}
\label{eq:newtrans}  
\end{align}

Then, $P$ can solve the new LP problem and distribute the solution share $\vec{y_i}^*$ to $M_i$, which securely reconstructs its solution share in the original problem with all the other parties.

\subsection{Secure Reconstruction}

Following the proof in \cite{HongIJER15,HongJCS12}, the optimal solution in the original problem $\forall i\in[1,n], \vec{x_i}^*$ can be reconstructed as below:

\begin{equation}
\forall i\in[1,n], \vec{x_i}^*=Q_{ii}\prod_{j=1,j\ne i}^nQ_{ij}\vec{y_i}^*
\end{equation}

As a result, all the parties $M_1,\dots, M_n$ should jointly reconstruct each solution share. Then, we present the secure communication protocol for the optimal solution reconstruction in Algorithm \ref{algm:re}.

\begin{algorithm}[!h]
	\small
	\begin{algorithmic}[1]
		\renewcommand{\algorithmicrequire}{\textbf{Input:}}
		\renewcommand{\algorithmicensure}{\textbf{Output:}}
		
		\FOR{$i \in [1,n]$} 
		\STATE $M_i$ generates a public/private key pair $(pk_i, sk_i)$ and sends the public key $pk_i$ to all the other parties $M_1,\dots, M_n$ 
		
		\STATE $M_i$ encrypts $\vec{y_i}^*$ with $pk_i$ to generate $Y=Enc_{pk}(\vec{y_i}^*)$, and sends $Y$ to the next party $M_j$ in Line 4
		
		\FOR{$\forall j\in [n,1], j\ne i, M_j$}

	\STATE $M_j$ updates $Y$ with $Q_{ij}$ (Line 6-8: $Y_a$ denotes the $a$th entry in $Y$)
	
	\FOR{each row $a$ of $Q_{ij}$}
	\STATE $M_j$ computes $Y_a \leftarrow \prod^{n}_{k=1} Y_k^{(Q_{ij})_{ak}}$
	\ENDFOR
		
		\STATE $M_j$ sends the updated $Y$ to the next party
		
		\ENDFOR
		
		\STATE the last party sends $Y$ to $M_i$
		
		\STATE $M_i$ decrypts $Y$ to obtain: $\prod_{j=1,j\ne i}^nQ_{ij}\vec{y_i}^*$
		
		\STATE $M_i$ reconstructs its share in the original optimal solution $\vec{x_i}^*=Q_{ii}\prod_{j=1,j\ne i}^nQ_{ij}\vec{y_i}^*$ (pre-multiplying by $Q_{ii}$)
		
		\ENDFOR
		
	\end{algorithmic}
	\caption{Secure Reconstruction}\label{algm:re}
\end{algorithm}

Finally, in the optimal energy sharing, each party $M_i$ can locally route the energy amount $x_{ij}^*\in \vec{x_i}^*$ to the recipient $M_j$ (note that $x_{ij}=0$ if $i=j$ holds).

\subsection{Privacy Preservation and Collusion Resistance}

\noindent\textbf{Privacy.} We now analyze the privacy leakage of the two protocols. For both extended secure transformation and secure reconstruction, there is no privacy leakage while executing the protocol under the definition of secure multiparty computation \cite{Yao86,Goldreich87} (all the messages received by all the parties can be simulated in polynomial time by repeating the protocols). Therefore, private inputs (e.g., demand, supply, and power quality of each party) can be protected. 

On the other hand, the information leakage in the outputs can be quantified:

\begin{itemize}
	\item The solver only learns the transformed optimization problem (the obfuscated shares of each party and the corresponding optimal solution).
	\item Each party only knows its share in the optimal solution, e.g., how much energy transmitted from itself to the energy recipient in the global optimal sharing.
\end{itemize}

\noindent\textbf{Handling Collusions.} The two protocols can also effectively handle potential collusions while solving the problem. None of those parties knows the actual overall transformation (aka. a combination of transformations), since each of $\{\forall i, \forall j, Q_{ij}\}$ is privately generated as a random nonnegative monomial matrix by $M_j$ (for transforming $M_i$'s shares). As a consequence, the solution reconstruction cannot be completed if any party $M_j$ is absent (missing $\forall i, M_{ij}$). Therefore, any number of microgrids (less than $n$) cannot collude with each other to infer private information from other honest microgrids while executing the protocol. The collusion resistant feature provided by the two protocols is equivalent to a trusted-third party. 

\section{Experiments}
\label{sec:exp}
\begin{figure}[!tbh]
	\centering
	\includegraphics[angle=0, width=1\linewidth]{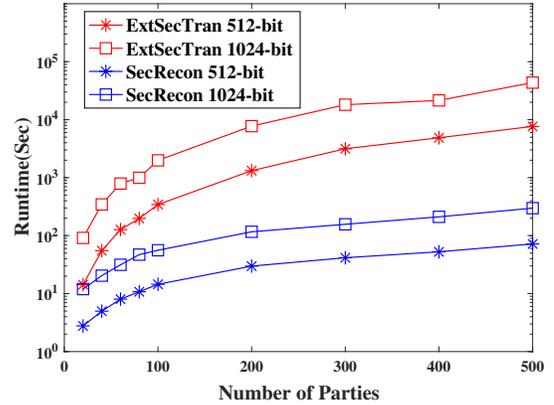}\vspace{-0.4in}
	\caption{Computational Costs} \label{fig:runtime}
\end{figure}

We have evaluated the performance of our revised secure transformation protocol and secure reconstruction protocol using two different key length (512-bit and 1024-bit) and varying number of parties (from 20 to 500). The computational costs of two protocols are plotted in Figure \ref{fig:runtime}. To significantly improve the security/privacy (provable), the protocols take longer time compare to \cite{HongIJER15}, and such computational costs are still tolerable with an polynomial increasing trend as the number of parties increase.

In addition, we present the communication overheads of the two protocols per party in Table \ref{tab:comm}. As the number of parties increase, the average bandwidth consumption (size of the transmitted messages) of the extended secure communication protocol and secure reconstruction protocol also grow polynomially.  Therefore, the two protocols can be implemented in most of the current networking environment. 

\begin{table}[!h]
	\small\caption{Communication Overheads} \centering
	\begin{tabular}{|c|c|c|}
		\hline
		Number of Parties& ExtSecTransform & SecReconstruction\\
		\hline
		20&0.00904 MB&0.0004 MB\\
		40&0.0761 MB&0.0019 MB\\
		60&0.261 MB&0.0045 MB\\
		80&0.624 MB&0.0078 MB\\
		100&1.23 MB&0.014 MB\\
		200&9.96 MB&0.051 MB\\
		300&33.5 MB&0.112 MB\\
		400&79.6 MB&0.119 MB\\
		500&155.6 MB&0.312 MB\\
		\hline
	\end{tabular}
	\label{tab:comm}
\end{table}

\section{Conclusion and Future work}
In this paper, we have extended the secure transformation and solution reconstruction in \cite{HongIJER15} to ensure provable security for securely solving the energy sharing optimization problem and implementing the optimal solution on the power transmission network. Novel secure communication protocols were proposed for all the parties to jointly transform their individual shares of the optimization problem, and jointly reconstruct their own shares in the optimal solution of the original problem. In the meanwhile, collusions can be handled with the secure communication protocols. In case that some parties disclose information to each other so as to learn other parties' private information, they cannot learn the actual transformation and reconstruction as long as at least one party is not colluding with them.

In the future, we will investigate other privacy preserving cooperative models among entities with local energy (viz. microgrids) on the power grid. For instance, global and local load balancing can be manipulated and further optimized via the cooperation among microgrids (e.g., scheduling \cite{HongEnergies17,liuscheduling17}). We intend to propose a privacy preserving cooperative model for them to jointly improving the global and local performance of the power generation, supply, storage and consumption.  

\subsection*{Acknowledgments}
This work is partially supported by the National Science Foundation under Grants No. CNS-1618221/1745894.

%\bibliography{ref} 
%\bibliographystyle{string,refs}
\bibliographystyle{abbrv}

\end{document}